\documentstyle[11pt]{article}
\renewcommand{\baselinestretch}{1.3}

\begin{document}
\newcount\nummer \nummer=0
\def\f#1{\global\advance\nummer by 1 \eqno{(\number\nummer)}
      \global\edef#1{(\number\nummer)}}

\newcommand{\pref}[1]{\ref{#1}}
\newcommand{\plabel}[1]{\label{#1}}
\newcommand{\prefeq}[1]{Gl.~(\ref{#1})}
\newcommand{\prefb}[1]{(\ref{#1})}
\newcommand{\prefapp}[1]{Appendix~\ref{#1}}
\newcommand{\plititem}[1]{\begin{zitat}{#1}\end{zitat}}
\newcommand{\pcite}[1]{\cite{#1}}
\newcommand{\plookup}[1]{\hoch{\ref{#1}}}

\let\oe=\o
\def\Di{\displaystyle}
\def\nn{\nonumber \\}
\def\be{\begin{equation}}
\def\ee{\end{equation}}
\def\ba{\begin{eqnarray}}
\def\ea{\end{eqnarray}}
\def\la{\plabel} \def\pl{\label}
\def\re{(\ref }
\def\rz#1 {(\ref{#1}) }   \def\ry#1 {(\ref{#1})}
\def\el#1 {\plabel{#1}\end{equation}}
\def\rp#1 {(\ref{#1}) }
\def\i{{\rm i}}
\let\a=\alpha \let\b=\beta \let\g=\gamma \let\d=\delta
\let\e=\varepsilon \let\ep=\epsilon \let\z=\zeta \let\h=\eta \let\th=\theta
\let\dh=\vartheta \let\k=\kappa \let\l=\lambda \let\m=\mu
\let\n=\nu \let\x=\xi \let\p=\pi \let\r=\rho \let\s=\sigma
\let\t=\tau \let\o=\omega \let\c=\chi \let\ps=\psi
\let\ph=\varphi \let\Ph=\phi \let\PH=\Phi \let\Ps=\Psi
\let\O=\Omega \let\S=\Sigma \let\P=\Pi \let\Th=\Theta
\let\L=\Lambda \let\G=\Gamma \let\D=\Delta

\def\wt{\widetilde}
\def\w{\wedge}
\def\0{\over } \def\1{\vec } \def\2{{1\over2}} \def\4{{1\over4}}
\def\5{\bar } \def\6{\partial }
\def\7#1{{#1}\llap{/}}
\def\8#1{{\textstyle{#1}}} \def\9#1{{\bf {#1}}}

\def\({\left(} \def\){\right)} \def\<{\langle } \def\>{\rangle }
\def\lb{\left\{} \def\rb{\right\}}
\let\lra=\leftrightarrow \let\LRA=\Leftrightarrow
\let\Ra=\Rightarrow \let\ra=\rightarrow
\def\ul{\underline}

\let\ap=\approx \let\eq=\equiv 
\let\ti=\tilde \let\bl=\biggl \let\br=\biggr
\let\bi=\choose \let\at=\atop \let\mat=\pmatrix
\def\CL{{\cal L}}\def\CX{{\cal X}}\def\CA{{\cal A}}
\def\CF{{\cal F}} \def\CD{{\cal D}} \def\rd{{\rm d}} 
\def\rD{{\rm D}} \def\CH{{\cal H}} \def\CT{{\cal T}} \def\CM{{\cal M}}
\def\CI{{\cal I}} \newcommand{\dR}{\mbox{{\sl I \hspace{-0.8em} R}}} 
  \newcommand{\dN}{\mbox{{\sl I \hspace{-0.8em} N}}}
\def\CP{{\cal P}}\def\CS{{\cal S}}\def\C{{\cal C}}

\begin{titlepage}
\renewcommand{\thefootnote}{\fnsymbol{footnote}}
\renewcommand{\baselinestretch}{1.3}
\hfill  TUW - 95 - 16\\
\medskip
\hfill  PITHA - 95/18\\
\medskip
\hfill  gr-qc/9508020\\

\begin{center}
{\LARGE {Classical and Quantum  Gravity in 1+1 Dimensions\\
Part I: A Unifying Approach}
 \\ \medskip  {}}
\medskip
\vfill

\renewcommand{\baselinestretch}{1} 
{\large {THOMAS 
KL\"OSCH\footnote{e-mail: kloesch@tph.tuwien.ac.at} \\
\medskip 
Institut f\"ur Theoretische Physik \\
Technische Universit\"at Wien\\
Wiedner Hauptstr. 8-10, A-1040 Vienna\\
Austria\\
\medskip 
\medskip THOMAS STROBL\footnote{e-mail:
tstrobl@physik.rwth-aachen.de} \\ \medskip
Institut f\"ur Theoretische Physik \\
RWTH-Aachen\\
Sommerfeldstr.\ 26--28, D52056 Aachen\\
Germany\\} }
\end{center}

\setcounter{footnote}{0}
\renewcommand{\baselinestretch}{1}                          

\begin{abstract}
  We provide a concise approach to generalized dilaton theories with
  and without torsion and coupling to  Yang-Mills fields.  Transformations
  on the space of fields are used to trivialize the field equations
  locally. In this way their solution becomes accessible within a few
  lines of calculation only.  In this first of a series of papers we set the
  stage for a thorough global investigation of classical and quantum
  aspects of more or less all available 2D gravity-Yang-Mills models.
\end{abstract}


\vfill
\noindent \hfill {\em Class.\ Quantum Grav.} {\bf 13} (1996), 965;\\
  \hspace*{\fill} Corr., {\em ibid} {\bf 14} (1997), 825.\\
\vfill
\end{titlepage}

\renewcommand{\baselinestretch}{1}
\small\normalsize

\section{Introduction}

One of the major open problems in theoretical physics is how to
construct a consistent theory of quantum gravity. This long-standing
issue has been approached in various different ways. 
Up to now it is not clear which of those approaches, ranging as far
as from non-perturbative canonical quantum gravity 
\pcite{Ashtekar} to string theory,  is most adequate. Also it may be that 
they all provide complementary, but legitimate viewpoints  \pcite{Au}.

In all of these approaches some control over the classical theory
seems indispensable.  In a path integral, e.g., the leading
contributions come from the local extremals of the action; or in the
canonical approach the space of observables is correlated directly to
the space of classical solutions modulo gauge transformations
(provided always that there are no anomalies). In addition to this the
relation between different approaches to quantum gravity might be
illuminated through considerations on the classical level. However, in
comparison to the vast infinity of possible solutions to the Einstein
equations only a negligible number of (exact) solutions is known and
the space of classical solutions (modulo gauge transformations) is
maybe even less seizable than the corresponding quantum theory.
 
The situation changes drastically, if one regards gravity models in
lower dimensions. Within the recent decade such models have attracted
considerable interest -- in three (cf., e.g., \pcite{Wit,Car}) as well
as in two spacetime dimensions (cf., e.g., \pcite{viele}). In the
present treatise we will restrict ourselves to two spacetime
dimensions with Minkowski signature. The claim of this and the
following papers is that in the case of Lagrangians describing
gravity without additional matter fields (a dilaton field is not
considered as matter in this context) there is complete control over
the space of classical solutions as well as on the space of quantum
states.  In Parts II,III \pcite{II} of the present series we will provide
a classification of all global solutions (modulo gauge
transformations) of more or less all available gravitational models in
1+1 dimensions without matter couplings! We allow for all
possible topologies of the spacetime manifold; and indeed there are
models with solutions with basically arbitrarily complicated topologies.
In Part IV \pcite{IV}, on the other hand, we will construct all
quantum states of the considered two-dimensional models in a
Hamiltonian approach \`a la Dirac.

What can we hope to learn from this? Is not the situation in two
dimensions just too far away from Einstein gravity in four
dimensions? In part this is true, certainly, but the point is that
there {\em are} questions which are not sensitive to the dimensional
simplification. One example is the issue of time (cf., e.g.,
\pcite{Isham}), which arises in any diffeomorphism invariant theory.

Another such question could be the interplay between a path
integral, a canonical Hamiltonian approach, and the topology of
spacetime. Irrespective of the dimension of
spacetime a Hamiltonian treatment implies a restriction to topologies
of the form $\S \times \dR$, where $\S$ is some (generically
space-like) hypersurface. (In two dimensions obviously $\S$ is either
a circle or a real line). In this way one excludes, e.g., the
interesting topic of topology changes. In a path integral, on the
other hand, there is no restriction to particular topologies. Now, in
two-dimensional Yang-Mills (YM) theories people invented some cut and paste
technique so as to infer transition amplitudes of non-trivial
spacetime topologies from the knowledge of the Hilbert space
associated to cylindrical spacetimes \pcite{YMWit}. It is realistic to
hope that something similar will be possible in the two-dimensional
gravity theories. The result might provide a first idea of what may be
expected of the  somewhat analogous issue in four-dimensional
canonical and path integral quantum gravity.

But there are also interesting technical issues that may be
investigated: One of these is the role which  degenerate metrics play
in a canonical framework, another one is the construction of an
inner product in a Dirac approach to a quantum theory,
a third one
a comparison of quantum theories for Minkowskian and Euclidean
signatures of the (quantized) metric. All of
these issues, taken up in Part IV of this work, are of current
interest in 4D quantum gravity \pcite{Ashtekar2}.

\medskip

The purpose of the present paper is to set the stage for a thorough
investigation of 1+1 gravity on the classical and quantum level. In
Section 2 we introduce the models under study. They comprise all
generalized dilaton theories \pcite{Banks}, where, in its reformulated
version \pcite{Kunst}, we allow also for nontrivial torsion
\pcite{LNP,Mod}.  In this way one captures theories such as
$R^2$-gravity with \pcite{KV} and without \pcite{R2} torsion or
spherically reduced 4D gravity \pcite{Hajetal}; but it should be
stressed that the class of considered theories is much more general.
Not included in the present treatise is axially reduced 4D gravity
\pcite{Nicolai}.
In Section 4 we further deal with dynamically
coupled Yang-Mills gauge fields. A coupling to fermion or scalar
matter fields (besides the dilaton), on the other hand, is mentioned
briefly only, but cf.\ also \pcite{CGHS,Zwie} for instance.

\hfill The breakthrough in the analysis of 2+1 gravity (with
Lagrangian\break
$\int d^3x  \sqrt{|\det g|} \,  R$) came about with its
identification  as a Chern-Simons gauge theory of the 2+1
dimensional Poincare group \pcite{Wit}. Similarly, in 1+1 
dimensions there are  two models that could be identified with 
standard gauge theories: The first of these is 
the Jackiw-Teitelboim (JT) model of 1+1 
deSitter gravity with Lagrangian \pcite{JT} \be
L^{JT}=-\frac12\int_M d^2x\sqrt{-\det g} \; \varphi \; (R-\L)\,
, \el JT where $R$ is the Levi-Civita curvature scalar of the
metric $g$ and $\varphi$ is a Lagrange multiplier field
enforcing the field equation $R = \L \equiv const$. Rewriting
this action in an Einstein-Cartan formulation, it is found to
coincide with a YM gauge theory of the BF-type
\pcite{Isler} (where the gauge group  is the universal covering
group of $SO(2,1)$ \pcite{versus}). Analogously, ordinary dilaton
theory \pcite{BH} may be reformulated as a BF-theory of the 1+1
dimensional Poincare group, if one promotes its cosmological
constant to a dynamical field which becomes constant on-shell
\pcite{Jackiw}. In all of these cases the respective group
structure greatly facilitated classical and especially quantum
considerations.

Within the class of theories considered in this paper the JT- and the
dilaton model are very particular (and comparatively simple). They
are, e.g., the only ones which, in a Hamiltonian formulation, allow
for global phase space coordinates such that the structure functions
in the constraint algebra become constants (cf.\ Part IV). Still some
of their features, such as a local degrees of freedom count, hold for
more complicated 2D gravity theories, too.  In a way the situation reminds 
one a bit of the Ashtekar formulation of 3+1 gravity with its similarities
but also its differences to a 4D YM theory. The question arises: 
 Given, say,  spherically reduced gravity or the already
somewhat more complicated $R^2$-theory, or even a generalized
dilaton theory defined by a potential function $V(\cdot)$ or
$W(\cdot,\cdot)$ (cf.\ Sec.\ 2 below): Is
there some kind of gauge theory formulation for them, which is similarly
helpful in the determination of the space of quantum states or
the space of classical solutions modulo gauge transformations?
In other words: Is there some unifying approach to all of
these 2D gravity theories generalizing in a nonlinear way  main features of
YM-type gauge theories?

Indeed, this question may be answered in the affirmative, as is
demonstrated at the beginning of Section 3. The key feature will be
the identification of Poisson brackets on an appropriate {\em target}
space associated with any generalized dilaton theory. The resulting
point of view not only allows for a unified treatment of
gravity-Yang-Mills systems in two dimensions, it also provides tools
for their classical and quantum analysis, which are hardly accessible
otherwise. This will be demonstrated first when employing the
formalism to solve the field equations in a particularly
efficient manner in the remainder of Section 3. There we will provide
the general local solution to the field equations of the general
model in the vicinity of arbitrary spacetime points. The extension of
this to the case of dynamically coupled YM-fields is
taken up in Section 4, finally. Section 4 includes also a brief 
summary of the results of this first part as well as a short outlook
on Parts II--IV. 

\section{Models of 1+1 Dimensional Gravity}

\la{Models}

\hfill It is well-known that in two dimensions the Einstein-Hilbert term,\\
$\int \sqrt{|\det g|} R d^2x$, does not provide a useful action for field
equations, as it is a `boundary term'. However, a natural approach to
find a gravity action in two dimensions is to dimensionally reduce the
four-dimensional Einstein-Hilbert action. Implementing, e.g.,
spherical symmetry,  by plugging
\be (ds^2)_{(4)}=g_{\mu \nu}(x^\m) dx^\m dx^\n -\Phi^2 \,
(d \vartheta^2 + sin^2\vartheta d \varphi^2),\quad \m, \n \in \{0,1\}, \ee
into the four-dimensional action and 
integrating  over the angle coordinates
$\vartheta$ and $\varphi$, 
one obtains the two-dimensional action \pcite{Hajetal}
\be L^{spher}=    \int_M d^2x \sqrt{-\det g} 
\, \,  \left[{1 \0 4} \Phi^2 R(g) + 
\2 g^{\m\n} \6_\m \Phi\6_\n \Phi  - {1 \0 2}\right]  . \el spher
Here  $R(g)$ denotes the  Ricci scalar of the 
Levi-Civita connection of the (two-dimensional) metric $g$ 
and $\det g \equiv \det g_{\m \n}$. 
Note that the field $\Phi(x^\m)$ is restricted to positive values by 
definition.\footnote{This 
could be avoided by introducing a new
  field variable proportional to $\ln \Phi$.} 
As a consistency check one may verify that  
the resulting field equations of this effective two-dimensional
action provide solutions to the four-dimensional
Einstein equations, and these are nothing but the Schwarzschild
solutions, parametrized,  according to
Birkhoffs theorem, by the Schwarzschild mass $m$:
\be g=\left(1- {2m \0 r}\right)(dt)^2 -\left(1- {2m \0 r}\right)^{-1}
(dr)^2 \,  .\el gspher 

The currently most popular action for a two-dimensional gravity theory is,    
however, the CGHS-model \pcite{CGHS} 
\ba 
 L^{CGHS}(g,\phi,f_i) & =&  L^{dil}(g,\phi) + L^{Mat}(f_i,g) \, \, , 
\la{CGHS} \\
 L^{dil}(g,\phi)& =& \int_M d^2x \sqrt{-\det g}  \, \exp (-2 \phi) \, \,
 \left[R+4 g^{\m\n} \6_\m \phi\6_\n \phi -\L \right]  \la{Dil} \\
  L^{Mat}(f_i,g) &=& \int_M d^2x \sqrt{-\det g} \,\sum_{i=1}^N \, 
  g^{\m\n} \6_\m f_i \6_\n f_i  
 \, \, .  \la{Mat}   
\ea
The first part of this action, $L^{dil}$, is the so-called `string inspired'
or dilaton  gravity action  \pcite{BH} ($\phi$ corresponds to the dilaton
field in string theory), 
the second part, $L^{Mat}$,  is the standard kinetic term for $N$
scalar fields $f_i$. The vacuum solutions ($f_i \equiv 0$) of
\re{CGHS})  are  of a similar form as \re{gspher}) (with identical
Penrose diagrams). Moreover, the classical
model can be solved  completely also when the $f_i$ are present
\pcite{CGHS}. The CGHS-model thus opened
the possibility  to discuss, e.g.,  the Hawking effect \pcite{Hawking} 
in a simplified two-dimensional framework \pcite{BHreview}. Also, motivated by
the classical solvability of this model, one may hope for 
an exact quantum treatment of \rz CGHS \pcite{Zwie}, allowing to test the
semiclassical considerations leading to the Hawking effect.  

By means of the field redefinition $\Phi :=  2 \sqrt{2}
\exp(- \phi), \, \Phi >0$,
the gravitational or dilaton part of \rz CGHS may be put into the form 
\be  L^{dil}(g,\Phi)   = 
   \int_M d^2x \sqrt{-\det g} \, \, \left[\mbox{${1 \0 8}$} 
\Phi^2 R+ \mbox{$\2$} g^{\m\n}  \6_\m \Phi\6_\n \Phi -
\mbox{${1 \0    8}$} \Phi^2 \L \right]   \, . \el Dil2
Now \rz Dil and \rz spher are found to show much similarity. 

Both of the examples given above are a special case of the general
action  
\be L^{gdil} (g, \Phi)= \int_M d^2x \sqrt{-\det g} \,\, \left[D(\Phi) R(g) + 
\mbox{$\2$} g^{\m\n}  \6_\m \Phi\6_\n \Phi - U(\Phi) \right]  
 \, . \el gDil
This action, which we will call generalized dilaton action,
was suggested first in \pcite{Banks}. It is the most general
diffeomorphism invariant action yielding second order differential 
equations for the metric $g$ and a scalar dilaton field
$\Phi$. In the following we will 
restrict ourselves to the case that $D$ has an  inverse function
$D^{-1}$   everywhere on its 
domain of definition. Also, for simplicity, we assume 
that $D$, $D^{-1}$, and $U$  are $C^\infty$. 

With these assumptions we may use\footnote{At this point the 
chosen nomenclature might appear bizarre. In the
sequel, however, $X^3$ will turn out to serve as the third target
space coordinate  of a  useful $\sigma$-model formulation of \re{gDil}).} 
\be X^3 := D(\Phi) \el X3
as a new field variable instead of $\Phi$.
Introducing \pcite{Kunst} instead of $g$ 
\be \wt g := \exp[ \rho(\Phi)] \, g \; \; , \;\;\; \rho = 
 \mbox{$\2$}  \int^\Phi {du \0 dD(u)/du}  + const \, \, ,
\el gtilde 
moreover, the action \rz gDil takes the simplified form
\be    L^{gdil} (\wt g,X^3) =  \int_M d^2x \sqrt{-\det \wt g} \,\, 
[X^3 R(\wt g) -V(X^3)] \,\, .\el Ltilde
Here we have put 
$V(z) := (U /  \exp  \rho)\left(D^{-1}(z)\right)$ and  
different constants chosen for the definition of $\rho$ rescale only  
the potential $V$.\footnote{Mainly we  use only one  symbol for a 
 function or functional, if  it is represented in
  different coordinates  (cf., e.g., \re{Dil}) and \re{Dil2})). 
To avoid misinterpretations we haven't done so in the case of $V$.} 
Note that this field-dependent conformal transformation allowed to get rid
of the kinetic term for the dilaton field.

In the case of $L^{dil}$ or $L^{CGHS}$, respectively, 
the transformation from $\phi$, $g$ to  $X^3$, $\wt g$ 
 proves specifically powerful \pcite{Ver}: With an appropriate
choice of $const$ in $\rho$ one obtains $\exp \rho(\Phi)=D(\Phi), 
\, U(\Phi) =\L D(\Phi)$ and thus $V(X^3)=\L=const.$ 
Since, moreover, $L^{Mat}$ is
invariant under conformal transformations, the action \rz CGHS becomes  
\be L^{CGHS}(\wt g, X^3, f_i) =  \int_M d^2x \sqrt{-\det \wt g} \,\, 
[X^3 R(\wt g) -\L] + L^{Mat}(f_i, \wt g) \, \, .
\el CGHStilde
In this formulation the classical solvability of the CGHS-model is 
most obvious: The variation with respect to $X^3$ yields $R(\wt
g)=0$. This implies that up to diffeomorphisms the metric $\wt g$
is Minkowskian. Thus the field equations resulting from the variation
with respect to the $f_i$ reduce to the ones of $N$ massless scalar fields
in Minkowski space. One then is left only to realize that due to the
diffeomorphism invariance only one of the three field equations $\delta
L^{CGHS}/\delta \wt g_{\m\n}(x) =0$ is independent \pcite{Sundermeyer} 
and that this one may be solved always for the Lagrange multiplier field
$X^3$ locally. Still, the
representation of $L^{CGHS}$ in the form \rz CGHStilde 
does not imply that the scalar fields 
$f_i$ and the original metric $g$ decouple completely. Rather one should
compare it to the introduction of normal coordinates for coupled harmonic
oscillators. To trace the coupling explicitly, one notices that the
transition from $\wt g$ to $g$ involves $X^3$, which in turn is coupled
directly (via $\wt g$ in \re{CGHStilde})) to the scalar fields $f_i$.

Let us represent $L^{gdil}$ in first order form. 
For this purpose we switch to the Cartan formulation
of a gravity theory, implementing the zero-torsion condition 
by means of Lagrange multiplier fields $X^\pm$: 
\be  L^{gdil} (e^a,\o,X^i) =  -2 \int_M  \,\, 
X_a De^a + X^3 d\o   + {V(X^3) \0 2} \, \varepsilon \,\,, \el PSX3
with 
\be \wt g =  2 e^-e^+  \equiv e^- \otimes e^+ + e^+ \otimes e^- 
\, \, \el tildevielbein
and  
\be De^a \equiv de^a + \varepsilon^a{}_b \o \wedge e^b  \, , \quad 
  a \in \{-,+\} \, ,  \qquad 
\varepsilon \equiv e^- \wedge e^+ \, \, ,
  \el torsion 
such that $e^\pm$ is the zweibein  in a light cone basis of the
frame bundle,  $\o$ (or $\o^a{}_b \equiv
\varepsilon^a{}_b \o$)  is the Lorentz or spin connection, 
and $\varepsilon_{-+}= +1$. Here we have used  $\varepsilon R=-2 d \o$.

We derived \rz PSX3 from the general action \rz gDil (for the case
that \rz X3 is a diffeomorphism). In this way the zweibein and spin
connection are interpreted as  quantities corresponding to the
auxiliary metric $\wt g$, which in turn is related to the `true'
metric via \re{gtilde}). In the following we will argue that \re{PSX3}),  
or its generalization 
\ba  L^{grav} &=&   \int_M  \,\, 
X_a De^a + X^3 d\o   + W((X)^2,X^3) \, \varepsilon \,\, , \la{grav}
\\
(X)^2 &\equiv& X_aX^a \equiv 2X^-X^+ \, , \ea
may be regarded also as a gravity theory with metric 
\be  g =  2 e^-e^+ \, \, .\el vielbein

First we note that $L^{grav}$ is invariant with respect to the
standard gravity symmetries, which are diffeomorphisms and local frame
rotations. Second, more or less by  construction the action \rz grav
is in first order form. It is not difficult to see then that the $X^i, \, i
= +,-,3$, are precisely the generalized momenta canonically conjugate to 
 the one-components
of the zweibein and the spin connection (the corresponding 
zero-components serve as
Lagrange multipliers for the constraints of the theory, cf.\ also 
Part IV). Obviously
there is a need
for momenta in any first order formulation of a gravity theory, so
the $X^i$ appear very natural from this point of view. 

Last but not least, \rz grav may be seen to yield further already
accepted  models of 2D gravity for some specific choices of the
potential $W$. Let us choose, e.g., $2W[(X)^2,X^3]=V(X^3)=
{(X^3)}^\g \L$ where $\L \neq 0$ and
$\g$ are some real constants. For $\g =1$ we  immediately recognize the good 
old Jackiw-Teitelboim model of two-dimensional deSitter gravity \pcite{JT}. 
For $\g \neq 1$, on the other hand, we may eliminate the field $X^3$  by means
of its equation of motion. Implementing, furthermore, the zero-torsion
constraint by hand again, the resulting Lagrangian is found to be of the form
$L \propto \int \sqrt{- \det g} R^{\g/(\g -1)}$. To obtain an integer exponent
$n$ for $R$, one sets $\g = n/(n-1)$. These are the purely geometrical 
Lagrangians for higher derivative gravity proposed in \pcite{R2}. 
So, e.g., the potential $W^{R^2}:= -(X^3)^2 +\L$ yields the Lagrangian 
\be L^{R^2}= \int_M d^2x \sqrt{-\det g} \, (R^2/16 + \L) \,  \el R2
  of two-dimensional $R^2$-gravity.

Similarly, the potential 
\be W^{KV}= - \a  (X)^2/2 -  (X^3)^2 +\L/\a^2 \el WKV
leads, upon elimination of the $X$-coordinates, to
\be  L^{KV} = \int [-{1 \0 4} d\o \w \ast d\o - {1\0 2\a} De^a \w \ast De_a
+ {\L \0 \a^2} \e] \, , \la{KV} \ee
proposed first in \pcite{KV}. 
This Lagrangian is the most general (diffeomorphism and frame 
invariant) Lagrangian yielding second
order differential equations for 
zweibein and spin connection. It is noteworthy that, in  contrast to
four dimensions \pcite{vanNieu}, it contains only three terms. 
Here one allowed for nontrivial torsion. 
All torsion-free theories  described by \rz grav have a potential 
$W$ that is independent of
$(X)^2$, or, equivalently, by Lagrangians of the form \re{gDil}) (with
$\wt g \rightarrow g$). 

Before we close this section, let us return to the case of spherical
symmetry \re{spher}). An appropriate choice of  
the integration constant in \re{gtilde}) yields  $g =  \wt g /\sqrt{X^3}$
and the potential $W$ becomes $W=V/2=1/4\sqrt{X^3}$ in this case. Thus, the
space of solutions to \rz spher will be reproduced from \rz grav with this
potential, if, according to \ry tildevielbein , $g := 2e^-e^+/\sqrt{X^3}$.
Alternatively, as we will find in the following section, the positive mass
solutions \re{gspher}) may be described also by \re{grav}) with potential
$W=1/(X^3)^2$, {\em if\/} we use the simpler identification \re{vielbein}),
$g:= 2e^-e^+$. 

In this section we have shown that \rz grav is a universal action 
for gravity theories in two dimensions. In the following section we will 
find it to be a special case of a $\sigma$-model defined by a Poisson
structure on a target space, an observation that  allows to solve
the theory in an elegant and efficient manner. 

\section{The Local Solutions of the Field Equations} 

\la{local}

With the notational convention \be A_- \equiv e_- \equiv e^+ \, ,
\,\,\, A_+ \equiv e_+ \equiv e^- \, , \,\,\, A_3 \equiv \o \, \el
identi we can rewrite the action \rz grav up to a boundary term as \be
L= \int_M A_i \wedge dX^i + \2 \CP^{ij}(X(x)) A_i \wedge A_j \, \el PS
with
\be \left(\CP^{ij}\right)(X) = \left( \begin{array}{lll}
0 & -W & -X^- \\
W & \phantom{-}0 & \phantom{-}X^+\\
X^- & -X^+ & \phantom{-}0 
\end{array} \right) \,\quad \, i,j
\in \{ -,+,3 \} \, , \el P where, as before, $W$ is a function of
$(X)^2\equiv 2X^-X^+$ and $X^3$. The first decisive observation is
that in this form the action is not only covariant with respect to
diffeomorphisms on the spacetime manifold $M$, but also with respect
to diffeomorphisms on the space of values of the fields $X^i$, i.e.\ 
on a `target space' $N=\dR^3$; we only have to define the
transformation of the $A_i$ and of $\CP^{ij}$ as those of one-forms
and bivectors on $N$, respectively. (The term $A_i \wedge dX^i$ in \rz
PS may then be interpreted as the pullback of a one-one-form $A=A_{\m
  i}dx^{\m} \wedge dX^i$ on $M \times N$ under the map of $M$ into the
space of fields and likewise the second term in \rz PS as the pullback
of the twofold contraction of $A$ with the two-tensor $\CP= (1/2)
\CP^{ij} \partial_i \wedge \partial_j$ on $N$, cf.\ 
\pcite{proce,Brief}).  The second decisive observation is that the
matrix $\CP$ obeys the following identity\footnote{The study of
  actions of the form \rz PS where $\CP^{ij}$ satisfies \rz Jacobi has
  been proposed also in \pcite{Ikeda}. However, the implications of
  the identity \ry Jacobi , recapitulated in what follows, have been
  realized only in \pcite{LNP,Mod,proce,Brief}.}
\be  { \partial \CP^{ij} \0 \partial X^l} \CP^{lk} + cycl.(ijk) =0 \,\,.
\el Jacobi
It establishes that $\CP$
is a Poisson structure on $N=\dR^3$. To see this one defines  
\be \{F,G\}_N = \CP^{ij}(X) { \6 F(X) \0 \6 X^i } 
{ \6 G(X) \0 \6 X^j }
\,  \el Poi
for any two functions $F$ and $G$ on $N$; the Jacobi identity for the 
Poisson brackets $\{ \cdot , \cdot \}_N$ on $N$ is then found to be 
equivalent to \re{Jacobi}). 
Vice versa, the Leibniz rule and antisymmetry of 
Poisson brackets ensures that they
can be written in the form \rz Poi with a skew-symmetric bivector $\CP$
on $N$. Thus we see that \re{PS}), and therefore also \re{grav}), may
be interpreted as a $\sigma$-model, where the world sheet $M$ is 
the spacetime manifold  and the target space $N$, which in the present case
equals $\dR^3$, is a Poisson manifold \pcite{Mad,Buchneu}.  

Note, however, that in our case the Poisson tensor \rz P  is
degenerate necessarily, 
as it is skew-symmetric and $N$ is three-dimensional here. Obviously at
points of $N$  where 
\be X^-=X^+=W=0 \, 
\el crit
$\CP$ has rank zero -- we will call these points `critical' further on
--, everywhere else it has rank two.  
Furthermore (as a consequence of the
Jacobi identity for  $\CP$) in the neighbourhood of generic (i.e.\
non-critical) points there  exists a  foliation
of $N$ into two-dimensional  submanifolds $\CS$, which are integral
manifolds of the set of Hamiltonian vectorfields (we will label them with
the coordinate function $\widetilde X^{1}$). 
Clearly, these leaves $\CS$ are symplectic (the restriction of $\CP$ onto
them is nondegenerate),
and one may choose coordinates $\wt X^2,\wt X^3$ such that on each of the
leaves $\CP|_{T^\ast \CS}$ (or its inverse) is in Darboux-form. In such an
adapted coordinate system $\wt X^i$, which we will call
Casimir-Darboux (CD) coordinate system\footnote{In some textbooks (cf., e.g.,
\cite{Buchneu}) such coordinates are called 
simply `Darboux coordinates'; however, we prefer the more suggestive term
above.}
$\CP \in \L^2(TN)$ takes the simple form
$\CP=  {\6 \0 \6 \widetilde X^2 } \wedge {\6 \0 \6 \wt X^3 }$. 

The notation of \rz PS allows to derive and depict 
the gravity field equations in a concise manner: 
\ba  dX^i +  \CP^{ij} A_j&=&0 \la{eom1a}\\
dA_i + \2 {\6 \CP^{lm} \0 \6 X^i} A_l \wedge A_m &=&0 \, . \la{eom1b} \ea
But what is more important, in order to solve these equations of
motion, the considerations above suggest the 
 use of  CD coordinates on the {\em target space} $N$. 
As $L$ is written in an $N$-covariant manner, the field
equations will still have the form \re{eom1a}, \ref{eom1b}),
only now $\CP$ is in Casimir-Darboux form. Explicitly this reads  
\ba d \wt X^1 =0 \; , && d A_{\wt 1}=0  \la{eom2a} \\
A_{\wt 2} = d \wt X^3 \; , && A_{\wt 3}
= -d \wt X^2 \; , \la{eom2b} \ea
while the remaining two field equations $d  A_{\wt 2}  =d  A_{\wt 3}=0$ 
are redundant obviously. In this form the solution of the
field equations  becomes a triviality:  
Locally \re{eom2a}) is equivalent to
\be \wt X^1= \mbox{const ,} \quad A_{\wt 1}= df \el Loesung
where $f$ is some arbitrary function on $M$, while \re{eom2b}) determines  
$A_{\wt 2}$ and $A_{\wt 3}$ in terms of the otherwise unrestricted 
functions $\wt X^2$, $\wt X^3$. 

Now we have to transform this solution back to the gravity variables
\rz identi only. Let us do this for the torsion-free case $W=V(X^3)/2$
first, Eqs.\ \re{PSX3}, \ref{Ltilde},\ref{gDil}). 
Here  
\be \wt X^i := \left({\2}\Big[(X)^2\!-\!\!\int\limits^{\;\;\;X^3}\!\! V(z)
dz\Big] \;,\; \ln|X^+| \;,\; X^3\right)  \el Xtildeneu
forms a CD coordinate system on $N$ on patches with $X^+ \neq 0$. 
To verify this we merely have to check $\{\wt X^1, \cdot \}_N = 0$ and 
$\{\wt X^2,\wt X^3 \}_N=1$, using the definition \re{P},\ref{Poi}) of
the brackets. 
 {}From $A_i= {\6 \wt X^j \0 \6 X^i} A_{\wt j}$  
we then  infer
\be e^+ \equiv A_- = X^+ A_{\wt 1} \, , \quad e^- \equiv A_+ = {1 \0
  X^+ }  A_{\wt 2} + X^-  A_{\wt 1} ,.   \el relationneu
By means of   \re{eom2b}, \ref{Loesung})  the metric $g=2e^+e^-$ thus 
becomes:
\be g=2 dX^3  df +  (X)^2 \, df df \, , 
\el gX3neu
where $(X)^2 = \int^{X^3} \, V(z) dz + const=: h(X^3)$ according to
\re{Xtildeneu}) and \re{Loesung}). Using $X^3$ and $f$ as coordinates
$x^0$ and $x^1$ on $M$, this may be rewritten as  \be g = 2
dx^0dx^1 + h(x^0) dx^1 dx^1 \el gh 
with  the function $h$ as defined above. In the case of $R^2$-gravity
\re{R2}), e.g., $h^{R^2}= -\mbox{${2\0 3}$} (x^0)^3 + 2 \L x^0 +C$,
where $C$ denotes the integration constant. In \re{gh}) $h$ depends
on one integration constant only,  which is a specific
function of the total mass (at least in cases where the latter may be
defined in a sensible way), reobtaining  what
has been called `generalized Birkhoff theorem' in various special
cases (cf.\ \pcite{R2,Katanaev}).

Maybe at this point it is worth mentioning that the coordinate
transformation \be r := x^0 \, , \;\; t := x^1 + \int^{x^0} {dz \0
  h(z) } \, , \el rtKoord well-defined wherever $h \neq 0$, brings the
generalized Eddington-Finkelstein form of the metric, Eq.\ \re{gh}),
into the `Schwarzschild form' \be g= h(r)(dt)^2 -{1 \0 h(r)} (dr)^2 \,
, \el gSSneu with the same function $h$. This confirms also that the
Schwarzschild case (with positive $m$) may be described by \re{grav})
also with the identification $g=2e^+e^-$: The potential $W=V/2 =
1/(X^3)^2$ yields $h(r)=C-1/r$, $C=const$, from which one finds
$m=C^{-(3/2)}$, after rescaling coordinates according to $r \to
\sqrt{C} \, r$ and $t \to t/\sqrt{C}$ ($C>0$).  In fact, we learn that
given {\em any\/} metric $g$ in (effectively) $1+1$ dimensions with (at
least) one Killing field, \rz grav with $g=2e^+e^-$ and $2W := V(X^3)
= h'(X^3)$ will provide an action which has $g$ within its space of
solutions.\footnote{This holds because any metric with a Killing field
  $v$ may be brought into the form \re{gh}) locally. Coordinate
  independently $h$ may be characterized as the norm squared of $v$ as
  a function of an affine parameter along a null-line, furthermore.
  For more details cf.\ Part II.} We remark, finally, that in the
present case of a torsion-free connection the Ricci scalar is just:
\be R = h''(x^0)\,. \el R
 
For a reformulated dilaton theory  \re{gDil}, \ref{Ltilde}) Eq.\ \rz
gh gives $\wt g$ only, which we will write as $\wt g=2d\wt x^0 dx^1 +
\wt h(\wt x^0)(dx^1)^2$ with $\wt h (\wt x^0) \equiv \int^{\wt x^0} \,
V(z) dz + const$. {}From Eqs.\ \re{X3},\ref{gtilde}) we have
$g=\exp[-\r\left(D^{-1}(\wt x^0)\right)]\wt g$. Here the coordinate 
transformation 
\be x^0(\wt x^0):=\int^{\wt x^0} \exp[-\r\left(D^{-1}(x)\right)] dx \ee
brings $g$ into  the form \re{gh}) again, where now  
$h(x^0)=e^{-\r(\wt x^0(x^0))}\wt h(\wt x^0(x^0))$. Let us illustrate
this by means of the dilaton theory \re{Dil2}). 
There $\wt h^{dil}(\wt x^0) = \L \wt x^0 + C$, $C$ denoting the
integration constant, and
$\wt x^0 = X^3 = D(\Phi)={1\0 8}\Phi^2=\exp(-2\phi) \in \dR^+$.
As  $\r=\ln \wt x^0$, Eq.\ \rz gtilde becomes $g= \wt g / \wt x^0$. 
The above coordinate transformation $x^0 = \ln\wt x^0$ 
yields 
\be h^{dil}(x^0) = \L + C \exp(-x^0) \el hDil 
with $x^0 \in \dR$. 
For \re{spher}) the analogous procedure  yields 
$h^{SS}(x^0) = 1 + 2C / x^0$,  leading  to the identification 
$m=-C$ in this case (cf.\ Eqs.\ \re{gSSneu}) and \re{gspher})).

The transition from $\wt g$ to $g$, although
conformal, may have important implications on the global structure of
the resulting theory, namely if due to a divergent conformal factor
the domain of $g$ is only part of the maximally extended domain of $\wt g$.
For instance, in the dilaton theory the maximal extension of $\wt g$ is
Minkowski space with its diamond-like Penrose-diagram, whereas the Penrose 
diagram of $g$, found by studying the universal coverings of the local charts
obtained above (cf., e.g., Part II), is of 
Schwarzschild-type. Although $X^3 = \wt x^0$ was defined for positive
values only, $\wt g$ 
 remains well-behaved for  $\wt x^0 \in \dR$
and may be extended to that values. The conformal factor in the
relation between $g$ and $\wt g$, $1/\wt x^0$, on the other hand,
 blows up at $\wt x^0 =0$. Correspondingly  $R(g)$ is seen to 
diverge at $\wt x^0 \equiv \exp x^0 = 0$, cf.\ Eqs.\
\re{R},\ref{hDil}) 
and  $g$ cannot be extended in the same way as $\wt g$. As a result 
the shapes of the Penrose diagrams are different. 


Such a behavior is generic also in the
case that \rz X3 maps  $\Phi \in \dR$ to a part of $\dR$ only, say,
e.g., to $(a,\infty)$ with an increasing $D$. The conformal 
exponent $\r$ \rz gtilde will diverge at  $\wt x^0=a$, as 
$D$ was required to be a diffeomorphism, and for potentials $U$ that do
not diverge too rapidly for $\phi \to - \infty$ the Penrose diagrams
of $\wt g$ and $g$ differ. 

\medskip

Let us now discuss the solution to the field equations for a general
potential $W$ in \re{grav}), 
using this opportunity to present also a more 
systematic construction of a CD-coordinate system. 

By definition a Casimir coordinate $\wt X^1$ is characterized 
by the equation $\CP(d \wt X^1, \cdot)=0$ $\Leftrightarrow 
(\6 \wt X^1 / \6 X^i) \CP^{ij}=0$. 
For $j=3$ the latter implies that $\wt X^1$ has to be a Lorentz 
invariant function of $X^\pm$, i.e.  
\be \wt X^1=\2 C\left[(X)^2,X^3\right] \el tildeX1
for some two-argument function $C=C(u,v)$.
For $j=\pm$ we then obtain
\be 2 W(u,v) C,_u + C,_v =0 \, \el CDiff
where the comma denotes differentiation with respect to the corresponding 
argument of $C$. As \rz CDiff is a first order differential equation,  it 
may be solved for any given potential $W$ locally, illustrating  
the general feature of a local foliation of Poisson manifolds for the special 
case  \ry P . An important  consequence of \rz CDiff is the relation $C,_u
\neq 0$. This follows as on the target space (!) we have $dC \neq 0$ 
(by definition of a target space coordinate function), and, according to 
Eq.\ \ry CDiff , $C,_u =0$ at some point implies that there 
also $C,_v=0$ and thus $dC=0$. Certainly, as $C,_u$ is a function on $N$, 
in contrast to $dC$ it  remains non-zero also upon restriction to the
submanifolds $C=const$. We  may verify this  also explicitly 
at the torsion-free example: \re{Xtildeneu}) yields $C,_u(u,v) \equiv 1$. 

Using the method of characteristics, \rz CDiff may be reduced to an ordinary
first order differential equation: We may express the lines of constant
values of the function $C$ in the form
\be {du \0 dv}=2W(u,v) \, .\el Diff
The constant of integration of this  equation is a
function of  $C$ in general; however, as clearly any function of a Casimir is
 a Casimir again, we may just identify the integration constant with $C$.
To illustrate this, we choose
\be 2W(u,v) := V(v) + T(v) u  \, \, . \el T 
We then obtain from \rz Diff
\be u = \left[ \int^v V(z) \exp \left(-\int^z T(y) dy \right) dz +
const(C) \right]
\, \exp \left(\int^v T(x) dx \right) \, , \el u
where the lower boundaries in the integrations over $T$ coincide.
Upon the choice $const(C) :=C$ \rz u gives
\be C (u,v) = u \, \exp \left(-\int^v T(x) dx \right) - 
\int^v V(z) \exp \left(-\int^z T(y) dy \right) dz   \, . \el C
The integrations on the right-hand side should be  understood as definite
integrals with somehow fixed, $C$-independent lower boundaries. Different
choices for these boundaries rescale $C$ linearly.

Let us specialize \rz C to some cases of particular interest. 
In the  case $T \equiv 0$, describing torsionless gravity and
discussed already above, it gives 
\be C= (X)^2 -  \int^{X^3} V(z) dz \, ,  \el CV 
in coincidence with the first entry of \re{Xtildeneu}). 
For the Katanaev-Volovich (KV) model of 2D gravity with torsion
\re{KV}), as a second example, \rz C and \rz WKV
yield upon appropriate choices for the constants of integration
and a rescaling by $\a^3$ 
\be C^{KV}= -2  \a^2 \exp (\a X^3) \left( W^{KV} +{2 X^3 \0 \a} -
{2 \0 \a^2} \right) \, . \el CKV
Here $C,_u (u,v) =  \a^3 \exp (\a v)$.

The remaining task is to find Darboux coordinates. On patches with either 
$X^+ \neq 0$ or $X^- \neq 0$ this is a triviality almost: According to 
the defining relations  \re{P}) of the Poisson brackets we have 
$\{X^\pm,X^3\}_N=\pm X^\pm$, so obviously $\pm \ln |X^\pm|$ and 
$X^3$ are conjugates. Altogether therefore 
\be \wt X^i := (\mbox{$\2$}C,\pm \ln|X^\pm|,X^3) \,  \el Xtilde
forms a CD coordinate system on regions of $N$ with $X^\pm \neq 0$, 
respectively. 

So, now we just have to repeat the steps following Eq.\ \re{Xtildeneu}) in
the more general setting of a potential $W(u,v)$. However, as we do
not want to restrict ourselves to potentials e.g.\ purely linear in
$u$, Eq.\ \re{T}), we do not know $C$ in explicit form. Still it is
nice to find that also in the case of a completely general Lagrangian
\re{grav}) the metric takes the form \re{gh}) locally. Moreover,
$h$ may be determined in terms of the Casimir function $C(u,v)$ and
the Killing field $\6/\6_1$ will be shown to be a symmetry direction
of all of the solution\footnote{This is not a triviality as, e.g., for
  nonvanishing torsion the connection $\o$ is not determined (up to
  Lorentz transformations) by the metric $g$ already.}.  For the sake
of brevity we display the calculation for both of the sets
\re{Xtilde}) of CD-coordinates simultaneously. This means that in the
following we restrict our attention to local solutions on $M$ that map
into regions of $N$ with $X^+ \neq 0$ {\em or} $X^- \neq 0$.

Inserting into the analogue of Eqs.\ \re{relationneu}) the solutions
\re{eom2b}, \ref{Loesung}) and re-expressing thereby the $\wt X^i$ in
terms of the original fields $X^i$ again, \re{Xtilde}), we obtain
(the upper/lower signs being valid for the charts $X^+$ resp.\ $X^-\neq 0$)

\begin{eqnarray}
  e^\pm \equiv A_\mp   &=& C,_u X^\pm df \nonumber \\
  e^\mp \equiv A_\pm   &=& \pm {1 \0 X^\pm } dX^3 + C,_u X^\mp df \nonumber \\
  \o \,\,\equiv A_3 \, &=& \mp d\ln |X^\pm| - C,_u W  df \, .
 \label{relation}
\end{eqnarray}
where in the last line we used \re{CDiff}).  
For the metric $g=2e^+e^-$
this yields 
\begin{equation}
   g=\pm 2C,_u \, dX^3  df + \bar h(X^3,C) \,  df df \,,
 \label{gX3}
\end{equation}
with \be \bar h(X^3,C) :=  C,_u{}^2 \cdot (X)^2 \; .\el hbar
\hfill Here $(X)^2$ is a function  of $X^3$ and $C$ by inverting
the field equation\\
$C((X)^2,X^3)=const$, whereas $C,_u$ in \rz hbar
and \rz gX3 is, more explicitly,  $C,_u \left((X)^2(X^3,C),X^3\right)$.
Note that according to the context $C$ 
either denotes a function of $X^i$ or the constant which it equals due to
the first equation \re{eom2a}). 
 
Now again we want to fix a gauge.  {}From \rz gX3 we learn that
 $C,_u dX^3 \wedge df \neq 0$, because otherwise  $\det g = 0$. This
 implies that we may choose $C,_u dX^3$ and $df$ as coordinate
 differentials on $M$. Let us therefore fix the diffeomorphism invariance
of the underlying gravity theory by setting 
\ba \int^{X^3} C,_u[(X)^2(z,C),z] dz &:=& x^0 \, , \la{x0} \\
 f &:=& \pm x^1 \, .\la{x1} \ea In this gauge $g$ is seen to take the form
\re{gh}) again with 
\be h(x^0) = \bar
h(X^3(x^0),C) \, , \el h where $X^3(x^0)$ denotes the inverse of
\re{x0}). 
The local Lorentz invariance may be fixed by means of 
 \be X^\pm := \pm 1 \, ,\el Lorentz
finally. Besides \re{Lorentz}, \ref{x0})
the complete set of fields then takes the form 
\begin{equation}
  e^\pm = C,_u dx^1 \, , \quad e^\mp = {dx^0 \0 C,_u} + \mbox{$\2$} (X)^2 \,
  C,_u dx^1 \, , \quad \o =  \mp W \, C,_u dx^1 \, ,
 \label{inKarte}
\end{equation}
and $X^\mp= \pm \mbox{$\2$} (X)^2$.
Here again $(X)^2$, $C,_u$, and $W$
depend on $X^3(x^0)$ and $C=const$ only, $C$ being the only
integration constant left in the local solutions. 

In the torsion-free case, considered already above, 
$C,_u=1$, $X^3(x^0)=x^0$, and, cf.\ Eqs.\ \re{hbar}, \ref{h}, \ref{CV}),
$(X)^2=h(x^0)=\int^{x^0}V(z)dz +C$, thus  reproducing the results
obtained there. For the KV-model \re{KV}), as an example for a theory
with  torsion, on the other hand, 
the above formulas yield 
$x^0 = \a^2 \exp (\a X^3)$, $x^0 \in \dR^+$, and 
\be h^{KV}(x^0) = {1 \0 \a}  \left\{C x^0 - 2 (x^0)^2 \, 
\left[(\ln x^0-1)^2+1- 
\Lambda\right] \right\} \,  \el hKV
for instance.  Certainly \rz R does not hold any more, instead we find
 $R = -{4 \0 \a} \ln \left({x^0 \0 \a^2} \right)$. Here this is obtained most
 easily by concluding $R = -4X^3$ from (the Hodge dual of) the three-component 
of \re{eom1b}). 

In the above we captured the solutions within regions of $M$ where
either $X^+ \neq 0$ or $X^-\neq 0$. Clearly, in regions where $(X)^2
=2X^+X^-\neq 0$ the two charts \re{inKarte}) must be related to each
other by a gauge transformation, i.e., up to a Lorentz transformation,
by a diffeomorphism. However, one of these two charts extends smoothly
into regions with only $X^+\neq 0$ (but possibly with zeros of $X^-$),
and $(+\leftrightarrow -)$ for the other chart.\footnote{In Part II
  regions with $(X)^2\neq 0$ will be called `sectors', while patches
  with merely $X^+\neq 0$ {\em or} $X^-\neq 0$, generically containing
  several sectors, will be our `building blocks' for the global extension.}
In this way the above
mentioned diffeomorphism may serve as a gluing diffeomorphism,
allowing to extend the generically just local solution \re{gh})
 to one that applies wherever $X^+$ and
$X^-$ do not vanish simultaneously.
Let us remark here, furthermore, that the two representatives
\re{inKarte}) are mapped into each other by \be e^+
\longleftrightarrow e^- \; ,\quad \omega \longleftrightarrow -\omega
\; ,\quad X^+ \longleftrightarrow -X^- \; ,\quad X^3
\longleftrightarrow X^3 \, .\label{5} \el Strafo This transformation
reverses the sign of the action integral \re{grav}) only and therefore
does not affect the equations of motion. {}From \re{5}) it is obvious
that the gluing diffeomorphism (cf.\ above) maps one set of null
lines onto the respective other one, leaving the form \re{gh}) of
$g=2e^+e^-$ unchanged. It corresponds to a discrete symmetry of
\re{gh}) (called `flip' in Part II), which is independent from the
continuous one generated by $\6/\6_1$. 
Further details shall be provided in Part II. 

We are left with finding the local shape of the solutions in the
vicinity of points on $M$ that map to $X^+=X^-=0$. Here we have to
distinguish between two qualitatively different cases: First,
$C=C_{crit}\equiv C(0,X^3_{crit})$, where $X^3_{crit}$ has been chosen
to denote the zeros of $W(0,X^3)$, and second, $C \neq C_{crit}$.
The special role of the critical values of $C$ is due to the fact that
the Poisson structure $\CP$ vanishes precisely at the points $X^+=X^-=0$
on the two-surfaces $C=C_{crit}$,  cf.\ Eq.\  \ry crit .

We start treating the non-critical case $C \neq C_{crit}$: For this,
one could attempt to construct a CD coordinate system valid in a
neighborhood of a (non-critical) point $X^+=X^-=0$.  At least in the
torsionless case this may be done in an explicit way, but while e.g.\ 
for ordinary dilaton gravity ($W\equiv const$) Darboux coordinates
are provided already by rescaling merely $X^+$ and $X^-$ (since
$\{X^+,X^-\}_N=W$), in the more general case the formulas are somewhat
cumbersome. We therefore follow a somewhat less systematic, but
simpler route here: {}From the three-component of \rz eom1a it is
straightforward to infer that points with $X^+=X^-=0$ are saddle
points of $X^3$. This suggests to replace the gauge conditions
\re{x0},\ref{x1}) by an ansatz of the form \ba \int^{X^3}
C,_u[(X)^2(z,C),z] dz &:=& x y + a \, , \la{x0neu}
\\ f &:=& b \ln x \, ,\la{x1neu} \ea where $a$ and $b$ are constants
to be determined below. As a first justification of
\re{x0neu},\ref{x1neu}) we find $C,_u dX^3 \wedge df$ to be finite and
non-vanishing on $(x,y) \in \dR^2$. Implementing the above conditions
in \ry gX3 , the metric becomes \be g=2b \, dx dy + b {2xy + b \, h(xy
  + a) \0 x^2 } \, (dx)^2 \, , \el saddle where $h$ is the function
defined in \ry h . For generic values of $a$ and $b$ \rz saddle is
singular at $x=0$. However, the choice \be a := h^{-1}(0) \qquad b:= -
2/h'\left(h^{-1}(0)\right) \el ab is seen readily to yield a smooth
$g$!

The singularity of the gauge choice \rz x1neu at $x=0$ was devised
such that it compensated precisely the singularity of the CD
coordinate system at $X^\pm=0$, used to derive \ry gX3 .
Indeed, the lines of vanishing $x$ or $y$
in \rz saddle may be seen to correspond to lines of 
vanishing $X^+$ and $X^-$, respectively. Also, they are Killing
horizons: According to \re{hbar}) zeros of $(X)^2$ coincide
precisely with the zeros of $h(x^0)$, indicating that the
Killing-field $\6/\6x^1$ (in charts \re{gh})) becomes null on those lines
(cf.\ also \ry gSSneu ).  
The charts \rz saddle provide a simple alternative to a generalized
Kruskal extension (cf.\ Part II). For Schwarzschild, $h(r)=1-2m/r$, it
is a global chart (as $h^{-1}(0)$ is single-valued), in the more
complicated Reissner-Nordstr\oe m case $h(r)=1-2m/r+q^2/r^2$ the
constant $a$ may take one of the two values $r_{\pm}=m \pm
\sqrt{m^2-q^2}$ and \rz saddle provides a local chart in the vicinity
of the respective value of $r=x^0$. A generalization of the right-hand
sides of \rz x0neu and \rz x1neu to $F(x) y + G(x)$ and $\int^x
dz/F(z)$, respectively, with appropriate functions $F$ and $G$, allows
even for global charts of (two-dimensional) spacetimes of the form
$\dR\, \times$ `null-lines'. For more details on this confer
\pcite{rnletter}, where also a more systematic approach to these
charts is presented, illustrating the considerations by Schwarzschild
and Reissner-Nordstr\oe m.

For \rz saddle to exist it is decisive that the corresponding zero $a$
of $h$ is simple, cf.\ the second Eq.\ \ry ab . For non-critical
values of $C$, e.g., all zeros of $h$ are simple.  This is
particularly obvious for torsionless theories, where $h=2X^+X^-$ and
$h'(X^3)=V(X^3)$, but holds also in general.  If $C \in \{C_{crit}\}$,
on the other hand, there exist zeros of $h$ of higher degree.  Then
the spacetime manifolds $M$ with varying $X^i$ do not contain the
critical points $X^i=(0,0,X^3_{crit})$.  This may be seen in two
different ways: First, studying extremals running towards such a
point, one finds the point to be infinitely far away, cf.\ Part II.
Second, from the field equations point of view: Taking successive
derivatives $\6/\6 x^\m$ of the Eqs.\ \re{eom1a}) and evaluating them
at the critical points, we find \be X^- \equiv 0 \, ,\;\; X^+ \equiv 0
\, ,\;\; X^3 \equiv X^3_{crit}=\hbox{const.} \el momdeSitter

\rz momdeSitter corresponds to additional, separate solutions of
the field equations not treated before. Actually, they come as
no surprise. More or less by definition the critical points \rz
crit  of  the target space constitute zero dimensional
symplectic leaves. It is a general feature of Poisson
$\s$-models, verified here explicitly in \rz momdeSitter and
the first Eq.\ of \re{Loesung}), that  the  image $X(x)$ of the
map from the worldsheet or space time $M$ into the target space
$N$  has to lie entirely within a symplectic leaf $\CS \subset
N$.

The remaining field equations \re{eom1b}), which are, more explicitly,
\be De^a =0 \, ,\,\, d\omega = -W,_v(0,X^3_{crit}) \e\,, \el deSitter
show that the solutions \rz momdeSitter have vanishing torsion and
constant curvature all over $M$.  The metric for such a solution can
be brought into the form \re{gh}), too, with $h(x^0)=
W,_v(0,X^3_{crit}) \cdot [(x^0)^2 +1]$.  This in turn determines the 
zweibein and spin connection up to Lorentz transformations.

\section{Summary and Extension to Gravity-Yang-Mills}

We demonstrated that any of the 2D gravity models introduced in
Section \ref{Models} is of Poisson $\s$-form \ry PS . Exploiting some
fundamental facts of Poisson structures, namely the local existence of
what we called Casimir-Darboux coordinates, the field equations
reduced to \re{eom2a},\ref{eom2b}), the solution of which is
immediate.  The relation between the original field variables and the
transformed ones, Eq.\ \ry relation , provided the general solution in
terms of the metric then.\footnote{The employed method may be viewed
  as a perfected generalization of what has been done previously
  in ordinary dilaton theory \pcite{Ver} or the KV-model \pcite{Solo};
  it is, however, not inspired by that works, but self-evident from
  the Poisson $\s$ point of view.  Let us note on this occasion that
  with {\em appropriate} gauge conditions the field equations of \rz
  grav may be solved in a maybe less elegant, but almost as
  straightforward manner, too \pcite{rnletter}
  (cf.\ also \cite{Kummerneu}).} With the choice of a
gauge the latter took the form \ry gh , where $h$ was parametrized by
a single meaningful constant (the value of the Casimir-function $C$).

\rz gh provided the local solution on strips with either $X^+ \neq 0$
or $X^- \neq 0$. For non-critical values of the Casimir constant $C$
(guaranteeing $h'|_{h=0} \neq 0$) the metric could be brought into the
form \re{saddle},\ref{ab}) in the vicinity of points $X^+=X^-=0$.  For
critical values of $C$, finally, we obtained the deSitter solutions
\rz deSitter in addition to the solutions \rz gh (which in this case
may not be extended to points of simultaneous zeros of $X^+$ and
$X^-$).

The Poisson $\s$-model formulation \rz PS of 2D gravity theories
provides the proper generalization of Yang-Mills (YM) gauge theories
advocated in the introduction.  Actually \rz PS is able to describe 2D
YM-theories with arbitrary gauge group. Identifying the target space
$N$ of a Poisson $\s$-model with (the dual of) some Lie algebra with
structure constants $f^{ij}{}_k$ and defining $P^{ij} := f^{ij}{}_k
X^k$, the action \rz PS is seen to become \be \int X^i F_i \el BF
after a partial integration, where $F= dA + A \wedge A$ is the
standard Lie algebra valued YM-curvature two-form. The local
symmetries of the BF-YM action \rz BF are the standard ones: $A \to
g^{-1} A g + g^{-1} d g, \, X \to g^{-1} X g$. The symmetries of the
general model \rz PS are a straightforward, nonlinear generalization
of this: \be \d_\ep X^i =
\ep_j(x) \CP^{ji} \, , \; \d_\ep A_i = d\ep_i + {\CP^{jk}}_{,i} A_j
\ep_k \, . \el symmetries These symmetries are the Lagrangian
analogues of what is generated by the constraints in a Hamiltonian
formulation (cf.\ Part IV). In the present context of \ry grav ,
where $N=\dR^3$ with Poisson bracket \ry P , the symmetry
transformations \rz symmetries are equivalent to diffeomorphisms and
local Lorentz transformations on-shell. This equivalence holds only
under the assumption of non-degenerate metrics $g=2A_+A_-$, however, a
feature shared also by the Ashtekar formulation of 3+1 gravity.  We
will see in Part IV (cf.\ also \pcite{versus}) how this seemingly
irrelevant restriction will lead to different factor spaces (even if
chosen representatives of gauge equivalence classes are restricted to
non-degenerate metrics $g$).

With the addition of one more term not spoiling the symmetries \rz
symmetries the model \rz PS is capable also of describing YM actions
$\int F \wedge \ast F$ \pcite{Mod,proce} or even G/G gauged WZW theories
\pcite{Anton} (cf.\ \pcite{Brief} for a pedagogical exposition). Of
more interest for our present intentions are, however, dynamically
coupled 2D gravity-YM-systems. Allowing for a dilaton- and
also $(X)^2$-dependent coupling constant $\a((X)^2,X^3)$, the action
for such a system has the form \ba L^{gravYM} &=& L^{grav} + L^{YM}
\label{GYM1} \\
 L^{YM}&=& \int {1\0 4 \a((X)^2,X^3)} tr (F \wedge \ast F) \, ,
\label{GYM2} \ea 
where the trace is taken in some matrix representation of the chosen
Lie algebra and $\ast$ is the Hodge dual operation with respect to the
dynamical metric $g=2e^+e^-$ of $L^{grav}$, Eq.\ \ry grav .  The
special case with an abelian YM-part (gauge group $U(1)$) and with
$\a$ and $W$ depending on $X^3$ only has received some attention in
\pcite{KunstU(1)} recently (but cf.\ also \pcite{Solo,PRD}). Let us
show in the following that the general combined system \rz GYM1 is of
Poisson $\s$-form again! As a byproduct many of the results of
\pcite{KunstU(1)} may be obtained as a lemma to the general theory of
Poisson $\s$-models (including  an exact Dirac quantization, cf.\ Part
IV as well as \pcite{PRD,LNP,Mod}). 

To begin with we bring  \rz GYM2 into
first order form: \be L^{YM} \sim {L^{YM}}'= 
\int \left( E^i F_i + \a((X)^2,X^3) E^i E_i \, 
\varepsilon \right) \, , \el equiYM
where the indices $i$ are raised and lowered by means of the Killing
metric and $\varepsilon \equiv e^- \wedge e^+$. 
The equivalence of $L^{YM}$ with ${L^{YM}}'$ is seen by integrating out
the `electrical' fields $E$ (either on the path integral level or just
by implementing the equations of motion for the $E_i$ back into the action,
in complete analogy to how we obtained \rz KV from \ry grav ). 
To avoid notational confusion let us rename 
$X^\pm, X^3$ into $\varphi^\pm, \phi$ and denote the YM-connection by
a small letter $a$.  
Then $L^{grav} + {L^{YM}}'$ reads
\be  {L^{gravYM}}'= \int \varphi^a De_a + \phi \,  d \o + E^j F_j +
\left[W((\varphi)^2,\phi)+ \a((\varphi)^2,\phi) \,  E^j E_j\right]
  \varepsilon \, , \el GYMprime
where $F_j \equiv d a_j + f^{kl}{}_j a_k \wedge a_l$ and
the indices $a$ run over $+$ and $-$ while the indices $j$ run from
1 to $n$, $n$ being the dimension of the chosen Lie group. 
After partial integrations (dropping the corresponding surface
terms) and the identifications 
\be X^i := (\varphi^a,\phi,E^j) \; , \quad A_i := (e_a,\o,a_j)\,, \el coor
\rz GYMprime becomes of Poisson $\s$-form \rz PS on an
$n+3$-dimensional target space with Poisson brackets:
\ba & \{\varphi^+,\varphi^-\}= W + \a E^j E_j \; , \quad
\{\varphi^\pm,\phi\}=\pm \varphi^\pm \; , & \nn
& \{\varphi^\pm,E^j\}= 
0=\{\phi,E^j\}\; ,  & \nn
& \{E^j,E^k\}=f^{jk}{}_lE^l \; . & \label{PoiGYM}  \ea

As any Poisson tensor also the one defined in \rz PoiGYM (note
$\CP^{ij} \equiv \{X^i,X^j\}$) allows for Casimir-Darboux coordinates
in the neighbourhood of generic points. If the rank of the chosen Lie
algebra is $r$ then the rank of the Poisson tensor is $n-r+2$ and there
will be $r+1$ such Casimir coordinates. Correspondingly there will
be $r+1$ field equations \rz eom2a and $n-r+2$ field equations \ry
eom2b . In the CD-coordinates the symmetries \rz symmetries take a
very simple form and it is a triviality to realize that again the
local solutions are parametrized by a number of integration constants
which coincides with the number of independent Casimir functions.
Note that in the present context \rz symmetries entail
diffeomorphisms, Lorentz transformations, and non-abelian gauge
transformations all at once. So, more or less without performing any
calculation, we obtain the result that the local solutions are
parametrized by $r+1$ constants now. In the case of \pcite{KunstU(1)}
$n=1, r=1$, and $r+1=2$. (Here we ignored additional exceptional
solutions of the type \ry momdeSitter , corresponding to maps into
lower dimensional symplectic leaves). 

The Poisson structure \rz PoiGYM has a very particular form: First the
$E^j$ span an $n$-dimensional Poisson submanifold of $N=\dR^{n+3}$.
Second, the Poisson brackets between the gravity coordinates
$(\varphi^a,\phi)$ and the coordinates $E^i$ of this submanifold
vanish. And, last but not least, the Poisson brackets between the
`gravity'-coordinates close among themselves up to a {\em Casimir}
function of the $E$-submanifold. With this observation it is a
triviality to infer the local form of $g$ of $L^{gravYM}$ from the
results of Section 3: On-shell $E^jE_j$ is some constant $C^{E}$. So
in all of the formulas of Section 3 we merely have to replace $W$ by
$W+C^{E} \a$. Thus the metric $g$ again takes the form \rz gh locally,
where now $h$ is parametrized by {\em two} constants $C^{E}$ and
$C^{grav}$ (`generalized Birkhoff theorem for 2D gravity-YM-systems').
For torsionless theories, e.g., \be h(x^0)=\int^{x^0}_c V(z) dz + \2
C^{E} \int^{x^0}_c \alpha(z) dz + C^{grav} \, , \ee where $c$ is some
fixed constant (and again $2W(u,v)=V(v)$).

The addition of a YM-part still rendered a theory with finitely many
`physical' degrees of freedom. There is at least one generalization of
the gravity action \rz grav that yields a (classically)
solvable theory with an {\em infinite} number of degrees of freedom.
This is obtained when coupling fermions of one chirality to \ry grav
\ \pcite{Heiko}, as was observed first in \pcite{FermionKummer} 
in the context of the KV-model \ry KV .

On the level of local considerations the models introduced in Section
\ref{Models} all look quite alike. They all have a one-parameter
family of solutions of the form \rz gh (in the neighbourhood of
generic points). As will be seen in Parts II,III of this work, this changes
drastically, if one turns to global considerations. The richness and
complexity of the global solutions is encoded in the kind in which the
target space $N=\dR^3$ foliates (more precisely `stratifies') into
symplectic leaves. E.g., a potential $V$ in \rz grav with many zeros
will lead to a topologically complicated stratification of $N$ into
two- and zero-dimensional symplectic submanifolds. Correspondingly, as
we will find in Part III, there will exist smooth solutions on
space-times $M$ with relatively complicated topologies. If, on the
other hand, $V$ has no zeros (such as in the JT- or the ordinary
dilaton theory), the foliation (stratification) of $N$ is quite simple
and the most complicated topologies of smooth space-time
solutions $M$  occuring are cylinders and M\"obius strips.

On the global level the solutions will be parametrized no longer by a
single parameter only (or, in the context of \ry GYM1 , no longer by
$r+1$ parameters).  Instead there will be $m+1$ (resp.\ $(m+1)$
times $(r+1)$) such continuous parameters labeling them (in
addition to further discrete parameters), where $m$ is the number of
independent non-contractible loops on $M$.  Here it should be noted
that given some particular model (via specifying $V$ or $W$) there
will be solutions with different topologies of $M$. Consequently the
solution space (defined as the space of all possible globally smooth
solutions modulo gauge transformations) of the chosen model will have
different components differing in dimension.\footnote{Not to be mixed
  up with the dimension of $M$, which certainly always remains two. It
  is the fundamental group of a possible $M$ which determines the
  dimension of the respective component of the solution space.} As at
present there is no classification of the solution space of a general
Poisson $\s$-model yet, in Parts II,III we will approach this issue in our
specific gravity context from a more traditional, pedestrian point of
view. First we will construct the universal coverings to the local
solutions studied in this paper. Thereafter we then investigate the
possible smooth factor solutions, keeping track of all arising
parameters. 

In Part IV, finally, we will turn to the quantum theory of \rz grav
or \rz GYM1 and compare the result to the classical solution space.
It is in this quantum regime where the full power of the Poisson
$\s$-formulation will become particularly apparent.

\section*{Acknowledgement}
We are grateful to  H.D.\ Conradi and W.\ Kummer  
for discussions and remarks concerning the manuscript. 
This work has been supported in part by the Austrian Fonds zur F\"orderung
der wissenschaftlichen Forschung (Project P10221-PHY).

\end{document}